\documentclass[5p,twocolumn,times,number]{elsarticle}

\usepackage{graphicx}
\usepackage{amsmath}   
\usepackage{subfigure}
\usepackage{textcomp}
\def\pmbanner{\vskip35pt\vskip35pt\vskip20pt}

\begin{document}

\begin{frontmatter}

\title{\pmbanner CMOS Monolithic Pixel Sensors based on the Column-Drain Architecture for the HL-LHC Upgrade}
\author[add1]{K. Moustakas\corref{cor}}
\ead{moustakas@physik.uni-bonn.de}
\author[add3]{M.Barbero}
\author[add2]{I. Berdalovic}
\author[add1]{C. Bespin}
\author[add3]{P.Breugnon}
\author[add1]{I. Caicedo}
\author[add2]{R. Cardella}
\author[add4]{Y. Degerli}
\author[add2]{N. Egidos Plaja}
\author[add3]{S. Godiot}
\author[add4]{F.Guilloux}
\author[add1]{T. Hemperek}
\author[add1]{T. Hirono}
\author[add1]{H. Kr\"uger}
\author[add2]{T. Kugathasan}
\author[add2]{C. A. Marin Tobon}
\author[add3]{P. Pangaud}
\author[add2]{H. Pernegger}
\author[add2]{P. Riedler}
\author[add1]{P. Rymaszewski}
\author[add2]{E. J. Schioppa}
\author[add2]{W. Snoeys}
\author[add3]{M. Vandenbroucke}
\author[add1]{T. Wang}
\author[add1]{N. Wermes}

\cortext[cor]{Corresponding author}

\address[add1]{Physikalisches Institut, Rheinische Friedrich-Wilhelms-Universität Bonn, Bonn, Germany}
\address[add2]{CERN Experimental Physics Department, CH-121 Geneve 23, Switzerland}
\address[add3]{Centre de physique des particules de Marseille, 163 Avenue de Luminy, Marseille, France}
\address[add4]{IRFU, CEA-Saclay, Gif-sur-Yvette Cedex, 91191 France}

\begin{abstract}
Depleted Monolithic Active Pixel Sensors (DMAPS) constitute a promising low cost alternative for the outer layers of the ATLAS experiment Inner Tracker (ITk). Realizations in modern, high resistivity CMOS technologies enhance their radiation tolerance by achieving substantial depletion of the sensing volume. Two  DMAPS prototypes that use the same "column-drain" readout architecture and are based on different sensor implementation concepts named LF-Monopix and TJ-Monopix have been developed for the High Luminosity upgrade of the Large Hardon Collider (HL-LHC) .\\LF-Monopix was fabricated in the LFoundry 150 nm technology and features pixel size of $50x250~\mu m^{2}$ and large collection electrode opted for high radiation tolerance. Detection efficiency up to 99\% has been measured after irradiation to $1\cdot10^{15}~n_{eq}/cm^{2}$. TJ-Monopix is a large scale $(1x2~cm^{2})$ prototype featuring pixels of $36x40~\mu m^{2}$ size. It was fabricated in a novel TowerJazz 180 nm modified process that enables full depletion of the sensitive layer, while employing a small collection electrode that is less sensitive to crosstalk. The resulting small sensor capacitance ($<=3~fF$) is exploited by a compact, low power front end optimized to meet the 25ns timing requirement. Measurement results demonstrate the sensor performance in terms of Equivalent Noise Charge (ENC) $\approx11e^{-}$, threshold $\approx300~e^-$, threshold dispersion $\approx30~e^-$ and total power consumption lower than $120~mW/cm^2$.

\end{abstract}


\end{frontmatter}

\section{Introduction}
Monolithic Active Pixel Sensors (MAPS) constitute an attractive alternative in high energy physics experiments as the building blocks of vertex detectors in high precision tracking applications. Their prominent advantage stems from the fact that the manufactured devices contain both the sensor and the front end electronics in the same silicon crystal and are ready to be used without the need for the expensive and labor-intensive process of fine pitch bump-bonding. Monolithic pixel sensors have already been successfully used in experiments with low radiation environments \cite{HFT_Greiner_2015,ALPIDE_Mager_2016}, but their radiation tolerance was limited. Advancements in CMOS imaging technologies that enable the use of high resistivity substrate and high voltage biasing can be exploited to achieve full depletion of the sensitive volume and generation of high electric fields to ensure fast charge collection by drift. During the phase-II of the HL-LHC upgrade, the ATLAS ITk will be improved \cite{ATLAS_upgrade} to cope with the unprecedented levels of radiation and ten times higher hit rate. A dedicated CMOS collaboration has been established to develop and characterize fully Depleted MAPS (DMAPS), for the outer layers of the ATLAS ITk. These devices are required to tolerate particle fluence up to $1.5\cdot10^{15}~n_{eq}/cm^{2}$ and a Total Ionizing Dose (TID) up to 50 Mrad. Moreover they must be able to comply with the 25ns timing requirement of the ATLAS experiment. DMAPS prototypes manufactured in different technologies have been reported with encouraging results regarding the sensor radiation tolerance  \cite{CCPD_Hirono_2016,Investigator_Pernegger_2017,LF_Hirono_2018,Wang_iworid}. Additionally, multiple nested wells offered by modern CMOS processes allow for complex readout circuitry to be implemented inside the pixel, enabling the use of fast readout architectures.

Two large scale DMAPS prototypes have been developed for the ATLAS ITk based on a common "column-drain" readout architecture derived from the FE-I3 front end chip \cite{FEI3)}. The simple in-pixel logic does not severely constrain the pixel dimensions and the hit rate capabilities of this approach are well established and exceed the requirements of the ITk outer layers. A different sensor implementation concept was pursued in each chip. The first prototype, called LF-Monopix, is based on a large collection electrode and is manufactured in the LFoundry 150 nm HV-CMOS process. The second prototype, called TJ-Monopix, employs a small collection electrode and is fabricated in the TowerJazz 180 nm imaging process using a novel modification to enhance radiation tolerance \cite{ModifiedProcess_Snoeys_2017}. The radiation hardness of LF-Monopix and the implemented large collection electrode sensor has been proven by beam test results. Full functionality of TJ-Monopix after irradiation is demonstrated for the first time and its analog performance is evaluated by laboratory and radioactive source measurement results reported in this work.

\section{Sensor Implementation}
\begin{figure}
	\begin{center}
		\subfigure[]{
			\includegraphics[width=0.325\textwidth]{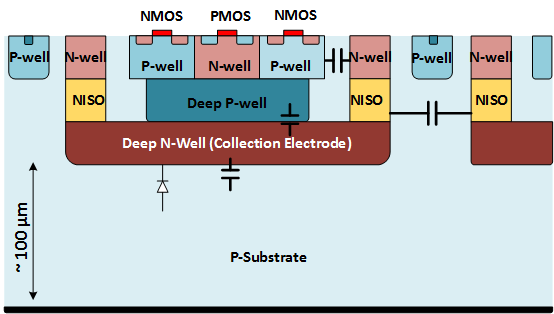}
			\label{fig:LF_Tech}
		}
		\qquad
		\subfigure[]{
			\includegraphics[width=0.33\textwidth]{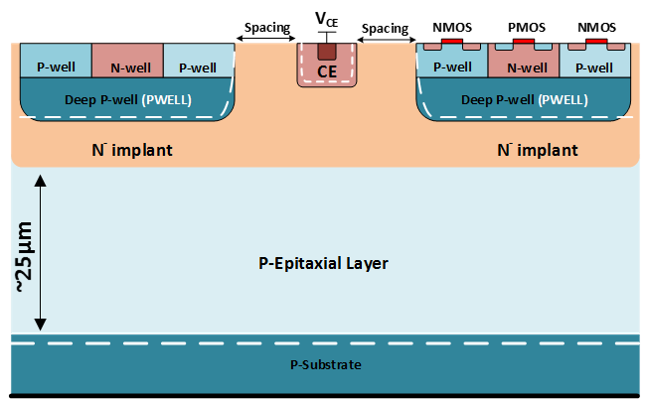}
			\label{fig:TJ_Tech}
		}
		\caption[]{\subref{fig:LF_Tech} LF-Monopix and \subref{fig:TJ_Tech} TJ-Monopix sensor design cross section} 
		\label{fig:Tech}
	\end{center}
\end{figure}

A crucial factor in the design of monolithic pixel sensors, that affects radiation tolerance and performance of the detector is the geometry of the collection electrode. In Figure~\ref{fig:LF_Tech} the cross section of the LF-Monopix pixel sensor is depicted. The collection electrode is formed by a large very deep n-well that includes all the in-pixel electronics. A high resistivity ($>$2~k$\Omega\cdot$cm) p-type substrate constitutes the sensitive volume that can be biased to high voltages. Its depth can vary from $750~\mu m$ to $100~\mu m$ after backside processing. The generated strong and uniform electric field ensures high radiation tolerance to Non Ionizing Energy Loss (NIEL) damage and high charge collection efficiency  \cite{CCPD_Hirono_2016,LF_Hirono_2018}. The disadvantage of this implementation is the large sensor capacitance ($\approx300-400~fF$) that degrades the analog performance in terms of noise and signal rise time, and has to be compensated with increased power consumption. Additionally, significant design efforts are needed to reduce the crosstalk noise from the in-pixel digital activity that is coupled to the input node.

In contrast, a different concept is pursued in TJ-Monopix. A small n-well is used as the collection electrode and the in-pixel electronics are implemented around it. The most important benefit of this geometry is the very small sensor capacitance ($\approx3fF$) that allows for very low noise and fast signal timing while keeping the power consumption minimal. Furthermore, crosstalk noise coupling is substantially reduced. 
The PMOS transistor n-wells are shielded by deep p-well implants. A 25$~\mu$m thick, high resistivity ($>$1~k$\Omega\cdot$cm) p-type epitaxial layer that is grown on a low resistivity substrate is used as the sensitive volume. To fully deplete the epitaxial layer and enhance radiation tolerance, a process modification has been developed \cite{ModifiedProcess_Snoeys_2017}. As depicted in Figure~\ref{fig:TJ_Tech}, a planar n- layer that realizes two p-n junctions is implanted over the full pixel area. As a consequence, the depletion boundary is extended laterally, to the otherwise undepleted part of the sensitive layer, even for small reverse bias voltages. Small pixel size is essential to reduce the distance between collection electrodes and ensure that the electric field is not degraded in the area between pixels. Measurement results from a test chip manufactured in the modified process to assess the sensor performance are very promising \cite{Investigator_Pernegger_2017}. Charge collection efficiency remains uniform and high ($>$95\%)  and the signal timing is still fast after irradiation to $1\cdot10^{15}~n_{eq}/cm^{2}$.

\section{Chip Design and Architecture}
\subsection {Chip Architecture}
\begin{figure}
	\centering
	\includegraphics[width=0.35\textwidth]{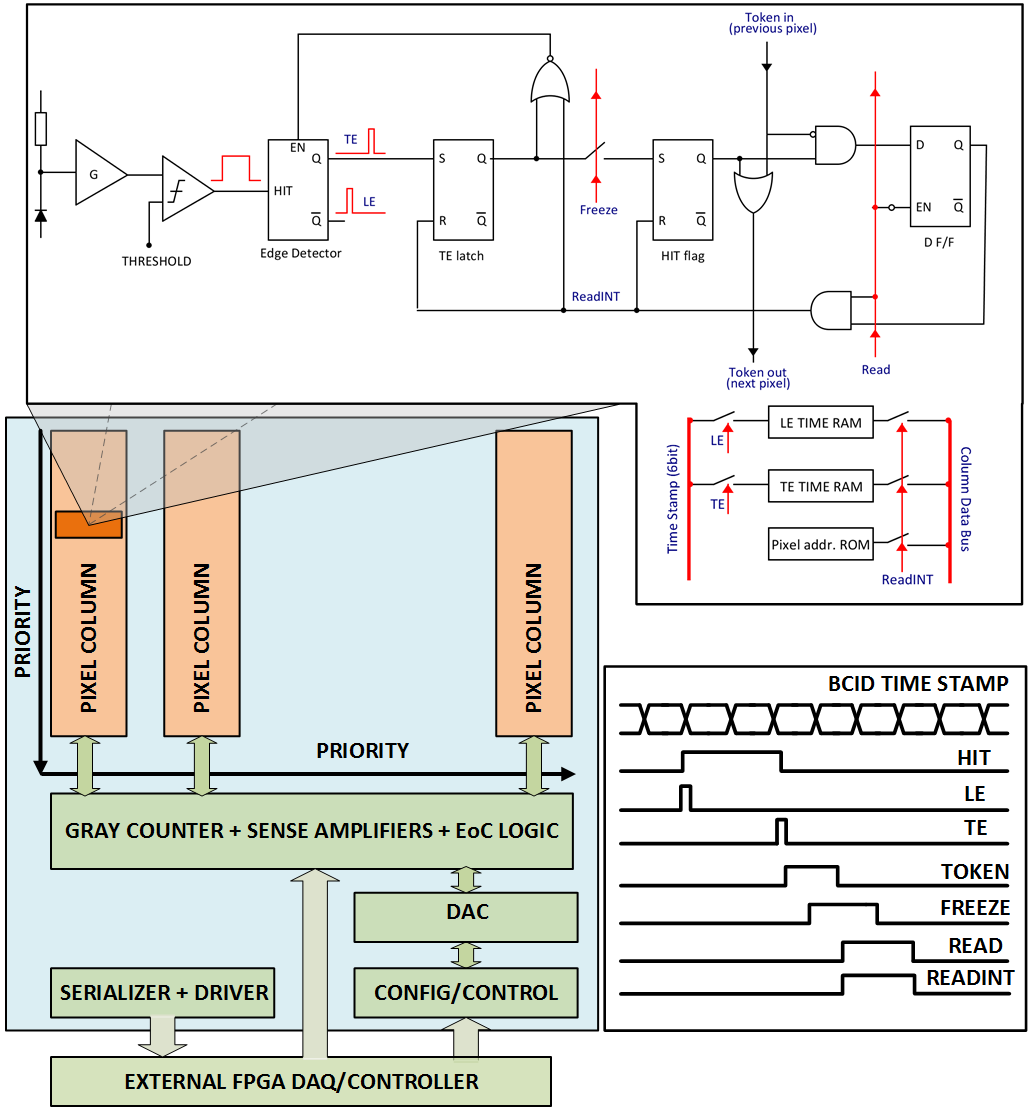}
	\caption{Chip architecture of LF-Monopix and TJ-Monopix} 
	\label{fig:chip_arch}
\end{figure}
Both prototypes follow the same basic architecture scheme that is depicted in Figure~\ref{fig:chip_arch}. Their differences in terms of readout logic are limited to implementation details and layout organization. A 40MHz Bunch Crossing ID (BCID) time stamp, generated by a gray counter is distributed over the whole matrix. After a hit pulse is produced by a pixel's analog front end, the Leading Edge (LE) and Trailing Edge (TE) information are stored in local in-pixel RAM memories. Arbitration over the common data bus is based on a token that is propagated with priority through the column from top to bottom and through different columns from left to right. A busy token flag is sent to the external Data Acquisition (DAQ) controller that initiates and controls the readout sequence. A local read signal is produced in the pixel that has been hit and has the highest priority to transmit the hit data. The data received from the RAM sense amplifiers at the end of each column are arbitrated by the end of column logic (EoC), and sent to the DAQ system using a 160 Mbps serial link. A trigger memory has not been implemented in the current prototypes and hit data is sent out immediately.

\subsection {LF-Monopix Chip Design}
The LF-Monopix chip \cite{LF_Hirono_2018,Wang_iworid,Wang_LFmono} is the successor to CCPD\_LF and LF-CPIX \cite{CCPD_Hirono_2016,LF_Hirono_2018} development line in 150 nm LFoundry technology and the first to incorporate a full standalone fast readout. It consists of a 129x36 matrix of pixels with 
$50x250~\mu m^{2}$ size, while its total size is $10x9.5~mm^2$. Special guard ring structures enable high voltage biasing up to 280 V before breakdown. An AC coupled charge sensitive amplifier (CSA) is used to amplify the input signal. The discriminator can be tuned by an in-pixel 4-bit DAC to reduce threshold dispersion. Each pixel consumes $\approx 36~mW$ of power yielding a total analog power consumption equal to $300~mW/cm^2$. The matrix is not uniform and consists of nine pixel flavors with different discriminator and CSA implementations. To reduce crosstalk noise, the peak transient currents of the readout logic and therefore the injected current to the sensitive node must be minimized. To this end, techniques that include current steering logic and column bus readout though a current-limiting source follower have been used \cite{Wang_LFmono}.

\subsection {TJ-Monopix Chip Design}
\begin{figure}
	\centering
	\includegraphics[width=0.36\textwidth]{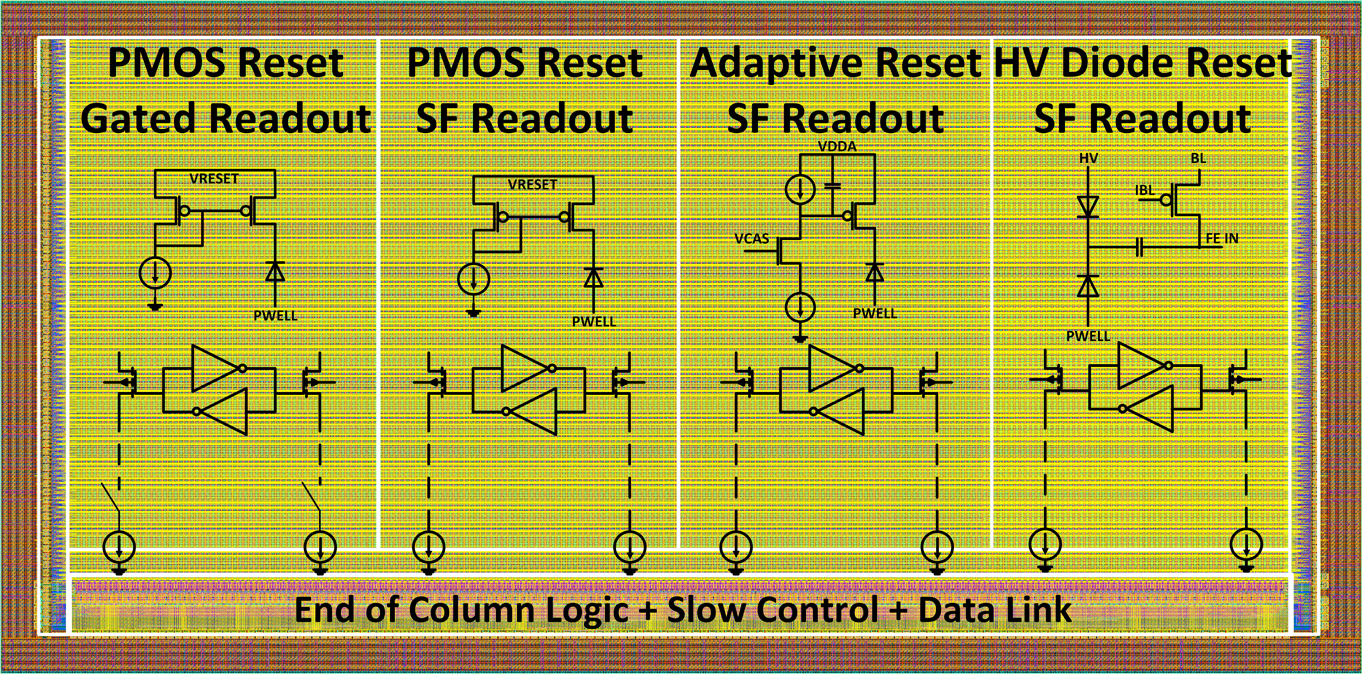}
	\caption{TJ-Monopix chip layout} 
	\label{fig:tjchip}
\end{figure}

Promising results of the modified process test chip encouraged the design of two large scale prototypes based on the same technology and front-end, called MALTA \cite{Berdalo_MALTA} and TJ-Monopix \cite{Wang_iworid,Me_NSS} featuring different readout schemes. The small sensor capacitance yields a very high input signal equal to $15~mV$ for the Most Probable Value (MPV) $\approx1500~e^-$. The high input signal to noise ratio can be exploited by a compact, non-conventional, low power front-end derived from the ALPIDE chip \cite{FE_Kim_2016}, that has been optimized for fast timing performance to meet the 25ns requirement.
TJ-Monopix chip layout is shown in Figure~\ref{fig:tjchip}. It features a 448x224 small size pixel ($36x40~\mu m^{2}$) matrix, while the total chip dimensions are $1x2~cm^2$. The analog power consumption is $<65~mW/cm^2$, and combined with the 6-bit time stamp distribution the total power consumption is only $120~mW/cm^2$. No in-pixel threshold tuning has been implemented since the front-end has been carefully designed for low threshold dispersion ($<30~e^-$). The matrix is split in four flavors. Apart from the standard pixel variation that features a PMOS input reset scheme, a novel leakage current compensation circuit and the possibility to bias the sensor with high voltage from the front-side are explored. A low power column bus readout variation is also implemented.

\section{Measurement Results}
\subsection{LF-Monopix}
\begin{figure}[!ht]
	\centering
	\includegraphics[width=0.32\textwidth]{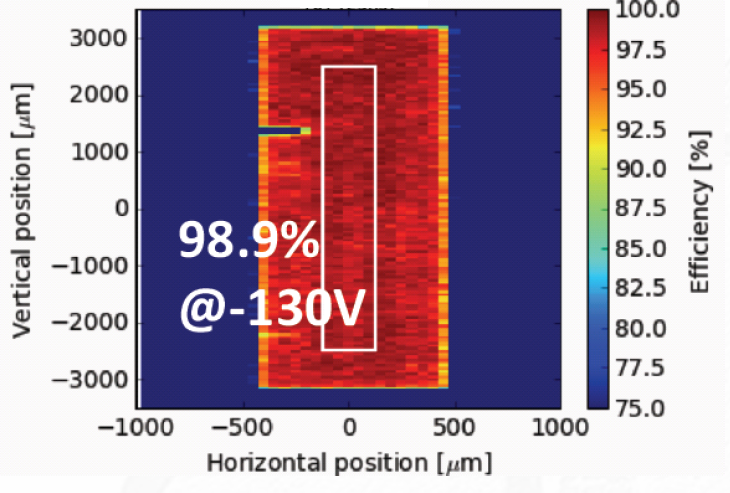}
	\caption{LF-Monopix detection efficiency measurement after irradiation to $1.5\cdot10^{15}~n_{eq}/cm^{2}$ \cite{LF_Hirono_2018}} 
	\label{fig:lfeff}
\end{figure}

LF-Monopix chips have been irradiated with neutrons up to $1.5\cdot10^{15}~n_{eq}/cm^{2}$ at the Jozef Stefan Institute reactor while the chips received an additional 1 Mrad background TID. The breakdown voltage has been measured equal to $280~V$ before irradiation \cite{Wang_iworid}, while after irradiation it drops to $\approx200~V$ and is still adequate to fully deplete the sensor. By applying a sophisticated threshold tuning algorithm based on the pixel noise occupancy, the threshold was tuned to $1500~e^-$ with a dispersion of $\approx100~e^-$. The achieved threshold value is low compared to the MPV, which is higher than 12 keV for an unirradiated sample and 4.5 keV after irradiation \cite{LF_Hirono_2018}. While the threshold value is not significantly affected after irradiation, the ENC did increase from $200~e^-$ to $350~e^-$ due to the background TID, with a dispersion equal to $30-70~e^-$ depending on the CSA flavor. The input charge to output voltage gain has been measured equal to $12~\mu V/e^-$ and is not degraded after irradiation. Figure~\ref{fig:lfeff} shows the high detection efficiency (98.9\%) of an irradiated LF-Monopix chip that was cooled down to $-40^o~C$, measured in an 2.5GeV electron beam \cite{LF_Hirono_2018}. The noise occupancy during the measurement was $<10^{-6}/25~ns$.

\subsection{TJ-Monopix}
\begin{figure}[!ht]
	\centering
	\includegraphics[width=0.34\textwidth]{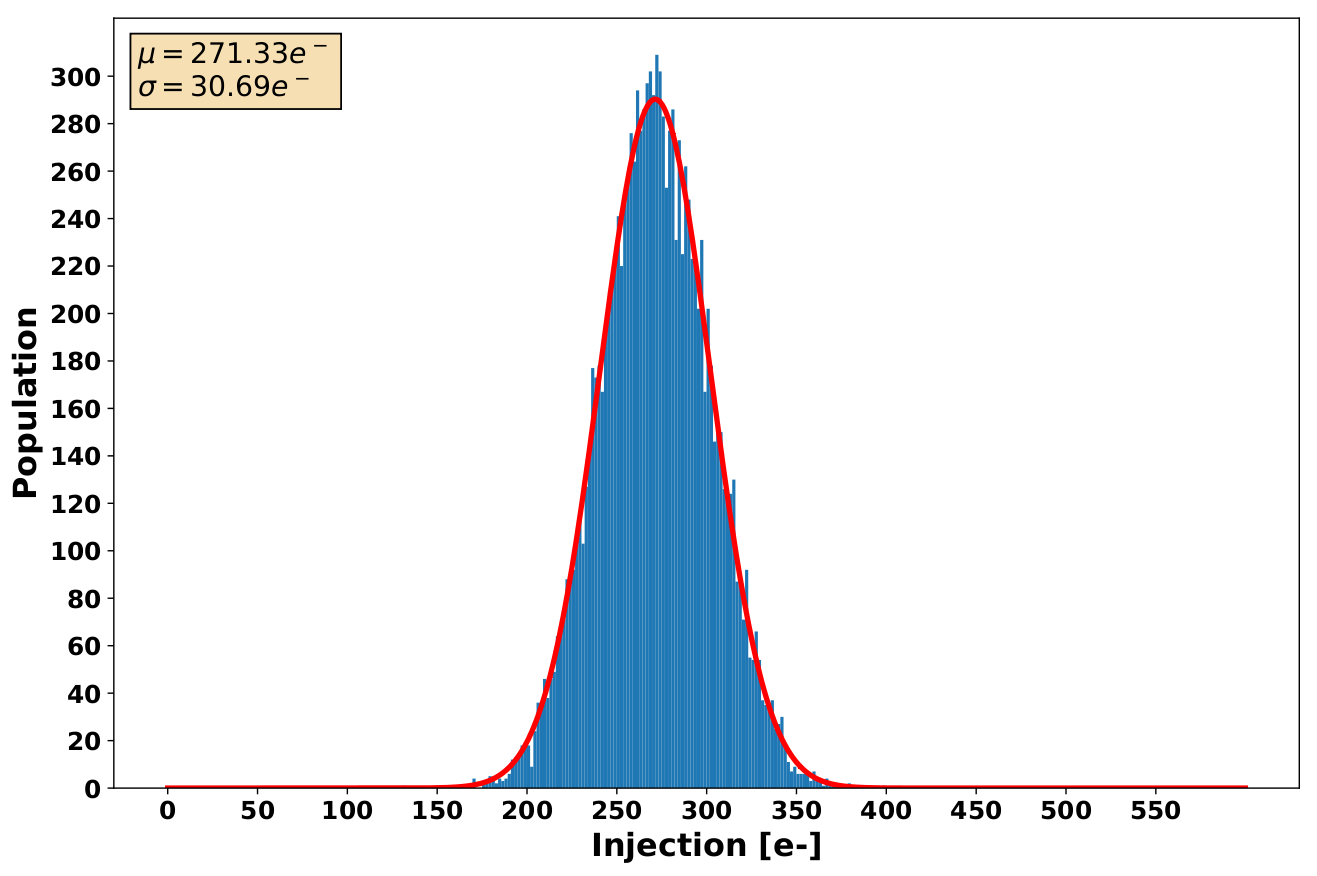}
	\caption{TJ-Monopix threshold scan: PMOS input reset flavor, p-well=-5V, substrate=-20V} 
	\label{fig:tjthr}
\end{figure}

While TJ-Monopix wafers were received in February 2018, and extensive characterization is currently in progress, first laboratory results are available. TJ-Monopix chips were also irradiated up to $1.5\cdot10^{15}~n_{eq}/cm^{2}$ and 1 Mrad backround TID, and remain fully functional. Since the sensor is fully depleted, the p-well potential inside the matrix can be separate from the epitaxial layer potential. The minimum p-well bias voltage is -6 V, while the substrate biasing is limited by the minimum voltage before punchthrough between the p-well and the epitaxial layer occurs and is equal to -20 V \cite{ModifiedProcess_Snoeys_2017}. The maximum front-side biasing voltage when using the special AC coupled flavor is +50V. The noise performance and threshold were measured for the PMOS input reset flavor with p-well biasing=-5 V and substrate biasing=-20 V. ENC was measured equal to $11~e^-$ with $0.8~e^-$ dispersion in agreement with simulation and the gain has been measured equal to $0.4~mV/~e^-$. The high gain and low noise are a direct consequence of the small sensor capacitance. As shown in Figure~\ref{fig:tjthr}, the threshold was measured equal to $271~e^-$ with dispersion equal to $30.7~e^-$. The noise occupancy that corresponds to these measurements is $\approx4\cdot10^{-8}/25~ns$. After irradiation, ENC is increased to $20~e^-$ and the threshold is increased to $470~e^-$ with dispersion equal to $50~e^-$.

\begin{figure}[!ht]
	\centering
	\includegraphics[width=0.34\textwidth]{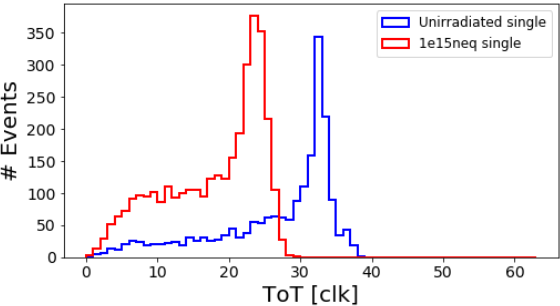}
	\caption{TJ-Monopix $^{55}Fe$ spectrum before and after irradiation to $1.5\cdot10^{15}~n_{eq}/cm^{2}$: HV-AC coupled pixel flavor, p-well=-0V, substrate=-16V, HV=+30V} 
	\label{fig:tjfe}
\end{figure}

Figure~\ref{fig:tjfe} depicts the spectrum of an $^{55}Fe$ source that was measured using the AC coupled HV flavor pixels and the analog ToT information. The bias voltages were set to p-well=0 V, substrate=-20 V, HV=+30 V. The measurement was repeated using an irradiated chip. The K$\alpha~(5.9 KeV)$ and K$\beta~(6.5 KeV)$ peaks produced by the electron capture decay of $^{55}Fe$ to $^{55}Mn$ are visible for the unirradiated case while for the irradiated chip, only the K$\alpha$ peak is clearly visible. It must be noted that the peak amplitude is degraded after irradiation due to the different front end settings, that were applied to increase the threshold and lower the noise occupancy.

\section{Conclusion}
Two large scale prototype chips based on the same fast readout architecture were implemented to demonstrate the feasibility of depleted monolithic active pixel sensors for the harsh radiation environment of the ATLAS ITk. The LF-Monopix radiation tolerance has been proven by means of $\approx99\%$ efficiency after irradiation to $1.5\cdot10^{15}~n_{eq}/cm^{2}$ while the TJ-Monopix chip remains functional after the same level of irradiation and first measurement results demonstrate the high analog performance of the small collection electrode implementation.

\section*{Acknowledgments}
This work is supported by the H2020 project AIDA-2020, GA no. 654168, and by the H2020 project STREAM, GA no. 675587


\end{document}